\documentclass[letterpaper, 10 pt, conference]{ieeeconf}  % Comment this line out if you need a4paper

\IEEEoverridecommandlockouts                              % This command is only needed if 
\usepackage{xcolor}

\usepackage{amsfonts}

\usepackage{graphicx}
\usepackage{tabularx}
\usepackage{tabulary}
\usepackage{makecell}

\usepackage{multirow}
\usepackage{subcaption}
\usepackage{booktabs}
\usepackage{threeparttable}

\let\labelindent\relax
\usepackage[inline, shortlabels]{enumitem}

\usepackage[natbib, bibencoding=utf8, citestyle=numeric, bibstyle=ieee, maxbibnames=999, maxcitenames=2, mincitenames=1, sortcites]{biblatex}
\bibliography{bibliography.bib}

\newcommand{\eg}{e.\,g., }
\newcommand{\ie}{i.\,e., }

\usepackage[acronym, shortcuts, nohypertypes={acronym}]{glossaries}
\newacronym{ACTH}{ACTH}{adrenocorticotrophic hormone}
\newacronym{eGeMAPS}{eGeMAPS}{extended Geneva minimalistic acoustic feature set}
\newacronym{HPA}{HPA}{hypothalamus–pituitary–adrenal}
\newacronym{FFNN}{FFNN}{feed-forward neural network}
\newacronym{GRU}{GRU}{gated recurrent unit}
\newacronym{MAE}{MAE}{mean absolute error}
\newacronym{PANAS}{PANAS}{\textit{Positive and Negative Affect Schedule}}
\newacronym{PASA}{PASA}{\textit{Positive Appraisal - Negative Appraisal}}
\newacronym{REG}{\textsc{REG-TSST}}{Regensburg Trier Social Stress Test}
\newacronym{SGD}{SGD}{stochastic gradient descent}
\newacronym[longplural={Stress Indices}]{SI}{SI}{stress index}
\newacronym{TSST}{TSST}{\textit{Trier Social Stress Test}}

\usepackage[capitalise, nameinlink, noabbrev]{cleveref}

\title{\LARGE \bf
Insights on Modelling Physiological, Appraisal, and \\Affective Indicators of Stress using Audio Features
}

\author{Andreas Triantafyllopoulos$^{1}$, Sandra Z\"{a}nkert$^{2}$, Alice Baird$^{3}$, \\ Julian Konzok$^{2}$, Brigitte M.\ Kudielka$^{2}$, and Björn W. Schuller$^{1,4}$% <-this % stops a space
\thanks{Author correspondence to: A.\,T., {andreas.triantafyllopoulos@uni-a.de}}% <-this % stops a space
\thanks{$^{1}$A.\,T. and B.\,W.\,S. are with EIHW -- the Chair of Embedded Intelligence for Health Care and Wellbeing, University of Augsburg, Germany}%
\thanks{$^{2}$S.\,Z., J.\,K., and B.\,M.\,K. are with the Institute of Psychology, University of Regensburg, Germany}%
\thanks{$^{3}$A.\,B. is with Hume AI, New York City, New York, USA}%
\thanks{$^{4}$B.\,W.\,S. is also with GLAM -- the Group for Language, Audio, \& Music, Imperial College, London, UK}%
\thanks{© 2022 IEEE.  Personal use of this material is permitted.  Permission from IEEE must be obtained for all other uses, in any current or future media, including reprinting/republishing this material for advertising or promotional purposes, creating new collective works, for resale or redistribution to servers or lists, or reuse of any copyrighted component of this work in other works.}%
}

\begin{document}
\maketitle
\thispagestyle{empty}
\pagestyle{empty}

%%%%%%%%%%%%%%%%%%%%%%%%%%%%%%%%%%%%%%%%%%%%%%%%%%%%%%%%%%%%%%%%%%%%
\begin{abstract}
Stress is a major threat to well-being that manifests in a variety of physiological and mental symptoms.
Utilising speech samples collected while the subject is undergoing an induced stress episode has recently shown promising results for the automatic characterisation of individual stress responses.
In this work, we introduce new findings that shed light onto whether speech signals are suited to model physiological biomarkers, as obtained via cortisol measurements, or self-assessed appraisal and affect measurements.
Our results show that different indicators impact acoustic features in a diverse way, but that their complimentary information can nevertheless be effectively harnessed by a multi-tasking architecture to improve prediction performance for all of them.
\end{abstract}

%%%%%%%%%%%%%%%%%%%%%%%%%%%%%%%%%%%%%%%%%%%%%%%%%%%%%%%%%%%%%%%%%%%%%%%%%%%%%%%%

\section{Introduction}
\label{sec:intro}
Stress is a major public health threat afflicting millions around the world and posing a risk to their physical and mental well-being.
A stress episode is characterised by the secretion of several hormones, of which cortisol is the most prominent~\citep{leistner2020hypothalamic}. 
% Cortisol is the end product of the \ac{HPA} axis, a hormonal axis which acts via the paraventricular nucleus (PVN) with the release of CRH and subsequent secretion of \ac{ACTH} from the pituitary into the blood, targeting the adrenal glands to release cortisol.
The release of such hormones is known to alter other physiological responses, \eg heart rate~\citep{gonulatecs2017analyzing}, which in turn affect speech, particularly during psychosocial stress~\citep{BRUGNERA201817}. 
Consequently, speech signals include valuable information on well-being~\citep{Cummins18-TII}, stress~\citep{Baird2019TSST, bairdevaluation}, states of emotional arousal~\citep{stappen2021muse}, and co-occurring conditions like anxiety~\citep{baird2020evaluation}.

While the activation of the \ac{HPA} axis and resulting hormonal response provide a satisfactory characterisation of stress from a psychoneuroendocrinological perspective, it is also desirable to characterise stress episodes from a psychological perspective~\citep{het2012stress}.
This adopts the viewpoint that stress can be thought of as
``a particular relationship between the person and the environment that is appraised by the person as taxing or exceeding his or her resources and endangering his or her well-being''~\citep{lazarus1984stress}.
Psychosocial stress can therefore be conceptualised as the result of a cognitive appraisal process resulting in an emotional, physiological, and behavioural stress response~\citep{gaab2005psychological}.
In lab conditions, this process can be accordingly measured through self-assessed questionnaires, such as the \ac{PASA} scale~\citep{gaab2005psychological}, administered before and after a stress episode.

Furthermore, perceived stress has been shown to correlate with fluctuations in self-assessed (negative) affect~\citep{watson1999panas, het2012stress}.
Self-assessment provides a useful tool for characterising stress episodes, namely, by quantifying how the subjects experience it from an emotional point of view.
This can also be measured pre- and post-stress using appropriate scales, such as the \ac{PANAS}, and serves as an alternative construct for stress characterisation~\citep{gaab2005psychological}.

Recently, the automatic detection of stress using various wearable sensors has seen increased popularity in the community, as timely detection of stress factors and corresponding reactions to them can facilitate the development of appropriate interventions to mitigate them.
Among others, electrodermal activity, heart-rate, and speech have been effectively used to detect and quantify stress responses.
As obtaining data from real-life stress episodes is obtrusive to the subjects' privacy, thus raising ethical concerns, studies typically monitor subjects in a controlled environment where stress is induced -- \eg in the well-known \ac{TSST}~\citep{kirschbaum1993trier}.
In our previous work, we investigated the feasibility of using speech obtained during a \ac{TSST} to predict cortisol measurements obtained at specified intervals before and after the test~\citep{bairdevaluation}.
Results from this study showed that audio-based features can be successfully used for that task, with recognition performance peaking for measurements obtained 10-20 minutes after the test (when individual cortisol measurements also peak).

However, predicting raw cortisol levels has limited usability in practical applications, where it is more important to measure the overall \emph{change} to cortisol levels resulting from a stress episode.
Moreover, a more holistic characterisation of stress responses is highly desirable.
The main goal of this work is to investigate the feasibility of using audio features to model different facets of stress episodes: a physiological indicator (cortisol), the subject's cognitive appraisal, and the subject's self-assessed affect.
For example, changes in self-assessed (negative) affect levels measured using the \ac{PANAS} scale prior to and following the \ac{TSST}, as well as changes to the \ac{SI} constructed based on the \ac{PASA} scale~\citep{gaab2005psychological}, provide alternative lenses with which to characterise stress episodes for each individual.
Specifically, we investigate three questions: a) to what extent each of these indicators can be modelled in isolation, b) how complimentary is the information contained in each of them, and c) whether there are subject-specific effects at play that influence how stress manifests in the audio signal.
% To the best of our knowledge, we are the first to investigate the automatic prediction of these stress indicators in parallel using speech signals.

The remainder of this paper is organised as follows.
In \cref{sec:data}, we describe the dataset, as well as the process we used to generate our targets.
\Cref{sec:setup} introduces the methodological approaches we used for modelling.
\Cref{sec:results} includes our results, and \cref{sec:conclusion} offers some concluding remarks.

%%%%%%%%%%%%%%%%%%%%%%%%%%%%%%%%%%%%%%%%%%%%%%%%%%%%%%%%%%%%%%%%%%%%%%%%%%%%%%%%%%%%%%
\section{Dataset}
\label{sec:data}

Our work is based on the \ac{REG} dataset, first introduced in~\citep{bairdevaluation}.
A total of 27 subjects (13 male, 14 female) aged [19-29] years old (median: 22) were recruited from the University of Regensburg campus and the surrounding community.
The entire study, including the questionnaires, instructions, and \ac{TSST} was conducted in German.
The study was approved by the Ethics Committe of the University of Regensburg on November 11, 2014 (No. 14-101-0283).

Subjects received verbal and written instructions upon arrival, followed by a resting period. 
During this time, a first saliva sample (T1 -- 30-45 minutes before the \ac{TSST}) was collected and a sugary drink (chilled herbal tea with 75\,g of glucose) was given to elevate blood glucose levels~\citep{zaenkert2020effect}.
The participants were given the \ac{PANAS} and \ac{PASA} questionnaires to fill in.
One minute before the next stage, another saliva sample was taken (T2 -- -1 minute). 
The subjects were then guided to the test room and introduced to the observers, at which point the \ac{TSST} commenced and the audio recording began.

Subjects were initially instructed to take the role of a job applicant and give a five-minute speech to present themselves as the best candidate for a vacant position.
At the end of the interview stage, subjects were given a mental arithmetic task lasting for a further five minutes, where they should serially subtract 17 from 2\,043 as quickly as possible.
After completion of both \ac{TSST} tasks, six more saliva samples were taken from the subjects (T3-T8).
The first was taken immediately after the end of the arithmetic task (T3 -- +1minute) and the rest at the following intervals after that: +10, +20, +30, +45, +60 minutes.
All saliva samples were analysed with a fluorescence-based immunoassay (DELFIA) to extract cortisol values measured in nanomoles per litre (nmol\,/l).
Additionally, the subjects were once again given the \ac{PANAS} and \ac{PASA} questionnaires to gauge their cognitive appraisal in retrospect and affect self-assessment.

Audio was captured using two different devices, namely a far-field microphone placed above the interviewee and a close-talk one worn by them (AKG PW45 presenter set).
For our experiments, we used the latter one as it resulted in cleaner recordings. 
All recordings were converted to $16$\,kHz, $16$\,bit, mono, WAV format and applying peak normalisation to $-1$\,dB for each audio file, \ie adjusting the loudness based on the maximum amplitude of the signal.

Based on the measured cortisol values and \ac{PASA}/\ac{PANAS} questionnaires, we constructed three targets corresponding to changes in each of the three indicators considered.
\begin{itemize}
    \item \textbf{Cortisol}: the first such target is the change in cortisol values with respect to the baseline caused by the \ac{TSST}.
    This is defined by the difference between the maximum cortisol value measured after (T3-T8) and the mean of the cortisol values measured before the \ac{TSST} (T1-T2).
    \item \textbf{Appraisal}: the second target is the difference between \acp{SI} measured before and after the \ac{TSST} using \ac{PASA}.
    \item \textbf{Affect}: the final target is the difference in negative affect before and after the \ac{TSST} measured with the \ac{PANAS} scale, as only negative affect has shown a correlation with stress~\citep{watson1999panas}.
\end{itemize}

\begin{figure}[t]
    \centering
    \includegraphics[width=.99\columnwidth]{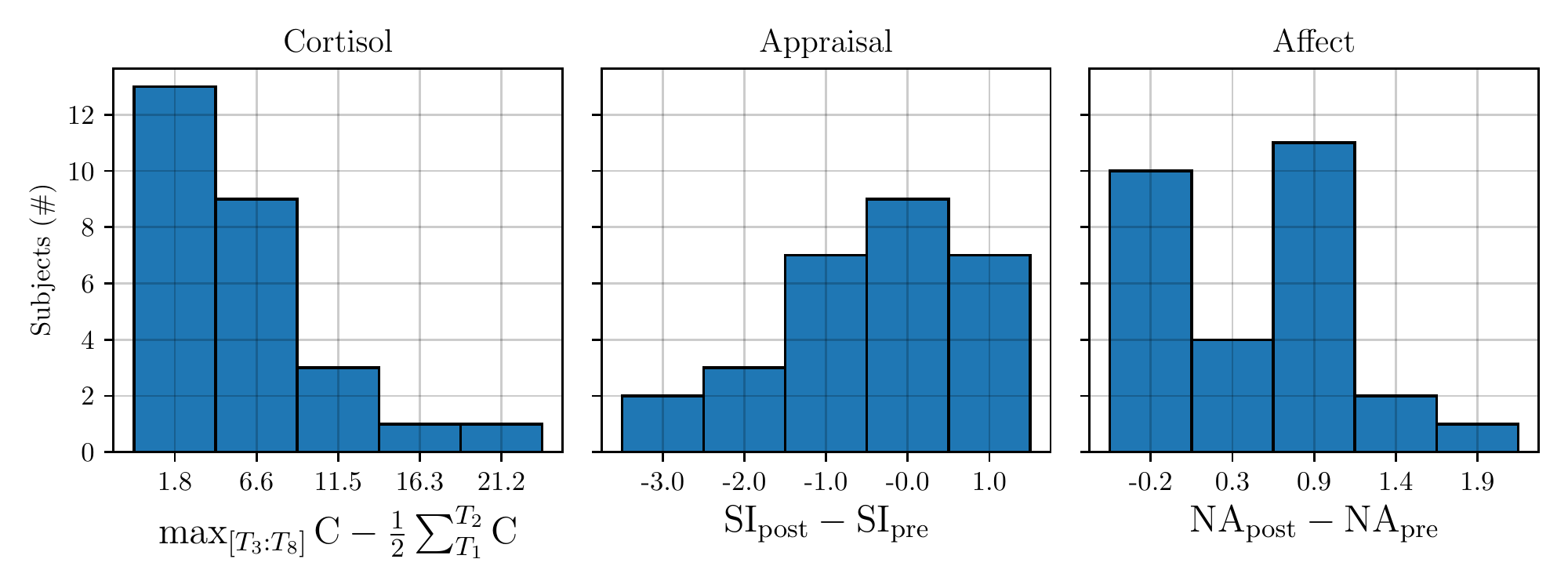}
    \caption{
    \textbf{(C)ortisol} (raw nmol\,/l values measured with DELFIA), appraisal (\textbf{(S)tress (I)ndices} constructed from the \ac{PASA} scale), and affect (\textbf{(N)egative (A)ffect} measured with the \ac{PANAS} scale) histogram plots.
    Constructs created as difference between post- and pre-stress values.
    }
    \label{fig:histograms}
\end{figure}
Histogram plots for the three targets are presented in \cref{fig:histograms}.
Most subjects show an increase in cortisol values (range: [-0.67, 23.6]\,nmol\,/l, median: 4.25 nmol\,/l) showing that their cortisol levels increased as a result of the \ac{TSST}.
Changes to \ac{SI} (range: [-3.5, 1.5], median: -0.38)
and negative affect (range: [-0.5, 2.2], median: 0.55) were more balanced on average.
For operationalisation purposes, all targets are scaled to a $[0-1]$ range when modelling.
This allows us to have a bounded target range, which makes training more stable, and additionally facilitates a coarse comparison between the results for each of the three targets.

% Taken together, those three constructs provide different lenses through which to interpret a stress episode.
% A key question is whether they all manifest in speech in similar ways; specifically, how each impacts speech features.
% Volume intensity (dB) is one speech feature known to vary during stress episodes~\citep{giddens2013vocal, bairdevaluation}.
% \Cref{fig:intensity} visualises the distribution of volume intensity for each of the three constructs.
% Although the targets are measured in continuous scales, we have binned them to three groups of equal size for better visualisation.
% Interestingly, the targets exhibit different patterns: cortisol and affect show high intensity values in their middle range where appraisal shows its lowest ones.
% Conversely, affect shows its lowest values for its low range where appraisal shows its highest ones.
% This inconsistencies are indicative of broader differences in the ways the three constructs manifest in the speech signal.
% \begin{figure}[t]
%     \centering
%     \includegraphics[width=.99\linewidth]{figures/intensity.pdf}
%     \caption{
%     Intensity scores in dB for the three targets investigated in our work: changes to cortisol levels (left), changes to \acpl{SI} measured by the \ac{PASA} scale (middle), and changes to self-reported negative affect measured by the \ac{PANAS} scale (right).
%     Targets binned to three percentile groups for visualisation purposes.
%     }
%     \label{fig:intensity}
% \end{figure}

% Alice stop here 

%%%%%%%%%%%%%%%%%%%%%%%%%%%%%%%%%%%%%%%%%%%%%%%%%%%%%%%%%%%%%%%%%%%%%%%%%%%%%%%%%%%%%%
\section{Experimental Setup}
\label{sec:setup}
Previous work has shown that segmenting the entire recording using overlapping windows, extracting segment-level features from each window, and subsequently treating the resulting sequence as input to a recurrent architecture, can be effectively used for stress recognition in a \ac{TSST} setting.
We follow the same modelling paradigm here.

Specifically, we use the openSMILE toolkit~\citep{eyben2010opensmile} to extract the \ac{eGeMAPS}~\citep{eyben2015geneva} with a window of $1$\,s and a hop size of $0.5$\,s.
This results in a total of $88$ features for each window; with the feature set comprising frequency- (pitch, jitter, formant frequency), energy- (shimmer, loudness, harmonics-to-noise ratio), and spectral-related parameters (alpha ratio, Hammarberg index, spectral slope, formant energy, harmonic difference).
As typically done in audio processing, those features are normalised to have $0$ mean and a standard deviation of $1$.
The normalisation parameters are computed on the training set and applied to the entire dataset.

The sequence of feature vectors is then fed into a \ac{GRU} network with $64$ hidden units, which sequentially process its input feature sequences and produces an equal number of hidden representations.
We used two variants: a vanilla one where the output of the \ac{GRU} is first mean-pooled and then fed into a linear layer to obtain the final prediction, and one where mean-pooling is substituted with attentional pooling.
In both cases, dropout with probability $0.2$ is applied to the output of the pooling layer for regularisation purposes.
All networks are trained for a maximum of $100$ epochs on the training set using \ac{SGD} with a learning rate of $0.001$ and a Nesterov momentum of $0.9$.
A batch size of $16$ is used during training, with all sequences padded to a maximum length of $1,200$ ($10$ minutes).
The models are early-stopped based on validation set performance.

Attentional pooling corresponds to a (dot-product) self-attention mechanism acting on the output of the \ac{GRU}.
Given a sequence $X\in\mathbb{R}^{\text{d}\times\text{T}}$, where $\text{d}$ is the feature dimensionality and $\text{T}$ the time dimension, dot-product self-attention uses a trainable weight matrix $W\in \mathbb{R}^{\text{d}\times1}$ to project that sequence into a time-sequence $Z\in\mathbb{R}^{\text{T}\times1}$.
This time-sequence is converted into a sequence of (pseudo-)probabilities using the softmax operator ($\alpha = \text{Softmax}(Z)$) which serve as weights to the input sequence $X$.
After the weights $\alpha$ are multiplied with $X$, the resulting sequence is summed over the time dimension to produce the output feature vector $F\in\mathbb{R}^{\text{d}}$.
Concretely:
\begin{align}
    Z &= X \times W \text{,} \\
    \alpha &= \text{Softmax}(Z)\label{eq:alpha} \text{, and} \\
     F &= \sum_1^T{(\alpha \cdot X)}\text{.}
\end{align}
The attention matrix $W$ then learns to assign importance to different points in the sequence $X$; a fact we later exploit to find which parts of the \ac{TSST} contribute more to the final decision.

Aside from the standard methodology described above, we investigate two additional facets of modelling the three herein considered stress indicators.
The first is whether the information contained in each of them is complimentary to one another.
An experimental investigation of that question can be provided by jointly modelling all three of them in a multi-task setup and measuring whether performance increases.
This is implemented by changing the output of the network to have a dimensionality of $3$.
All other settings remain the same.

The second question pertains to the extent in which the manifestation of stress in voice depends on the individual; individual differences would in turn manifest in the features we are extracting.
A typical way to mitigate such differences is to perform speaker-based feature normalisation: instead of computing global normalisation parameters based on the training data, and subsequently applying it to all instances, as is the standard process, the data from each speaker is normalised independently (including speakers in the dev/test set).
If such a normalisation were to improve performance, it would serve as an indication that individual effects play an important role.
Therefore, we conduct additional experiments using this form of normalisation.

Altogether this results in four configurations:
\begin{itemize}
    \item Vanilla \ac{GRU} using single- (\textbf{\ac{GRU}-STL}) and multi-task learning (\textbf{\ac{GRU}-MTL})
    \item Attention \ac{GRU} using single- (\textbf{A\ac{GRU}-STL}) and multi-task learning (\textbf{A\ac{GRU}-MTL})
\end{itemize}
each tested with both standard and speaker normalisation.

%%%%%%%%%%%%%%%%%%%%%%%%%%%%%%%%%%%%%%%%%%%%%%%%%%%%%%%%%%%%%%%%%%%%%%%%%%%%%%%%%%%%%%
\section{Results}
\label{sec:results}
\begin{table}[]
    \centering
    \caption{
    Cortisol, appraisal, and affect \acs{MAE} on the development and test set (dev/test).
    Targets were constructed as difference of values measured before and after the \ac{TSST}, and their scale was normalised to $[0-1]$.
    Cortisol corresponds to raw cortisol values (measured in nmol\,/l), appraisal to \acfp{SI} obtained via the \ac{PASA} scale, and affect to negative affect assessed with \ac{PANAS}.
    Test results only reported for the best-performing combination of each target to avoid overfitting (selection done on the dev set).
    }
    \label{table:results}
    \begin{tabular}{cc|ccc}
        \toprule
        \multirow{2}{*}{\thead{Model}} & \multirow{2}{*}{\thead{Normalisation}} & \multicolumn{3}{c}{\thead{Mean Absolute Error}}\\
         &  & \thead{Cortisol} & \thead{Appraisal} & \thead{Affect} \\
        \midrule
        \multirow{2}{*}{\thead{GRU-STL}} & Standard & .20/- & .33/- & .27/-\\
         & Speaker & .18/- & .31/- & .25/- \\
        \multirow{2}{*}{\thead{GRU-MTL}} & Standard & .16/- & .24/- & .26/-\\
         & Speaker & .15/- & .22/- & .26/- \\
        \multirow{2}{*}{\thead{AGRU-STL}} & Standard & .18/- & .29/- & .27/- \\
         & Speaker & .22/- & .23/- & .25/- \\
        \multirow{2}{*}{\thead{AGRU-MTL}} & Standard & .16/- & .27/- & \textbf{.24/.26} \\
         & Speaker & \textbf{.14/.19} & \textbf{.21/.26} & .28/- \\
        \bottomrule
        \end{tabular}
\end{table}
\begin{figure*}[!t]
    \centering
    \begin{subfigure}{0.48\textwidth}
    \includegraphics[width=\linewidth]{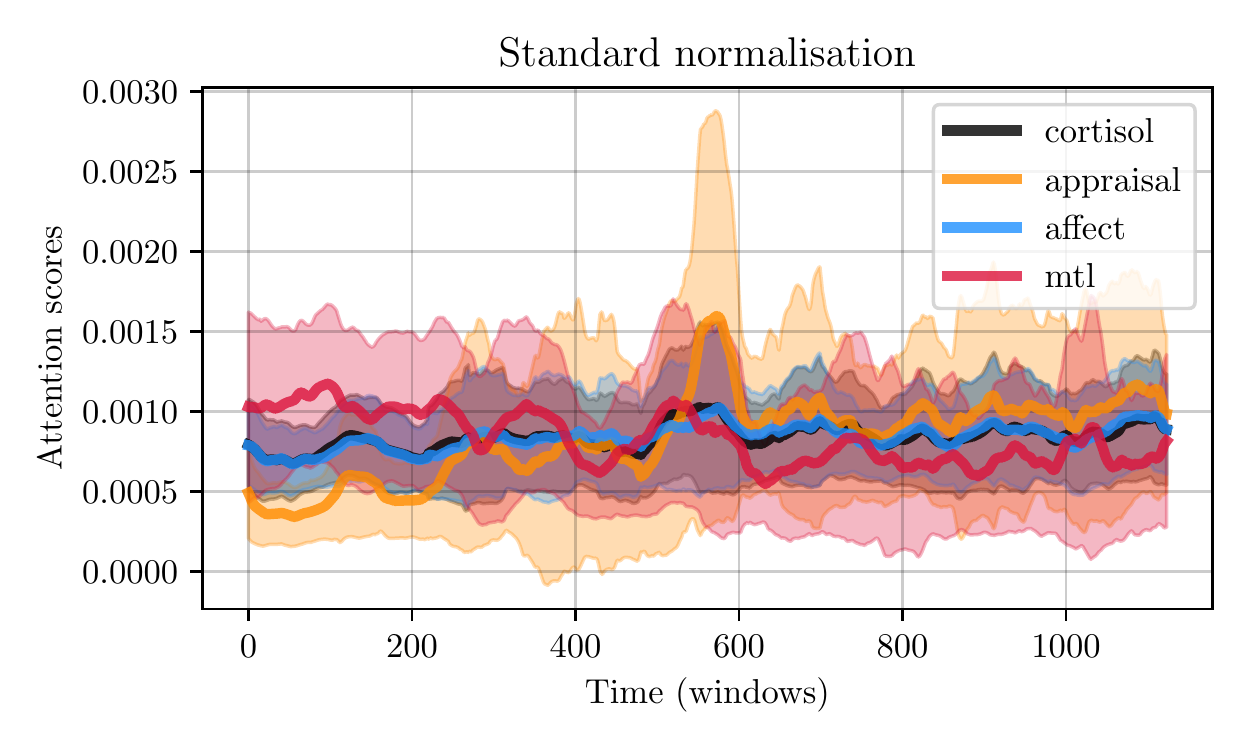}
    \end{subfigure}
    \begin{subfigure}{0.48\textwidth}
    \includegraphics[width=\linewidth]{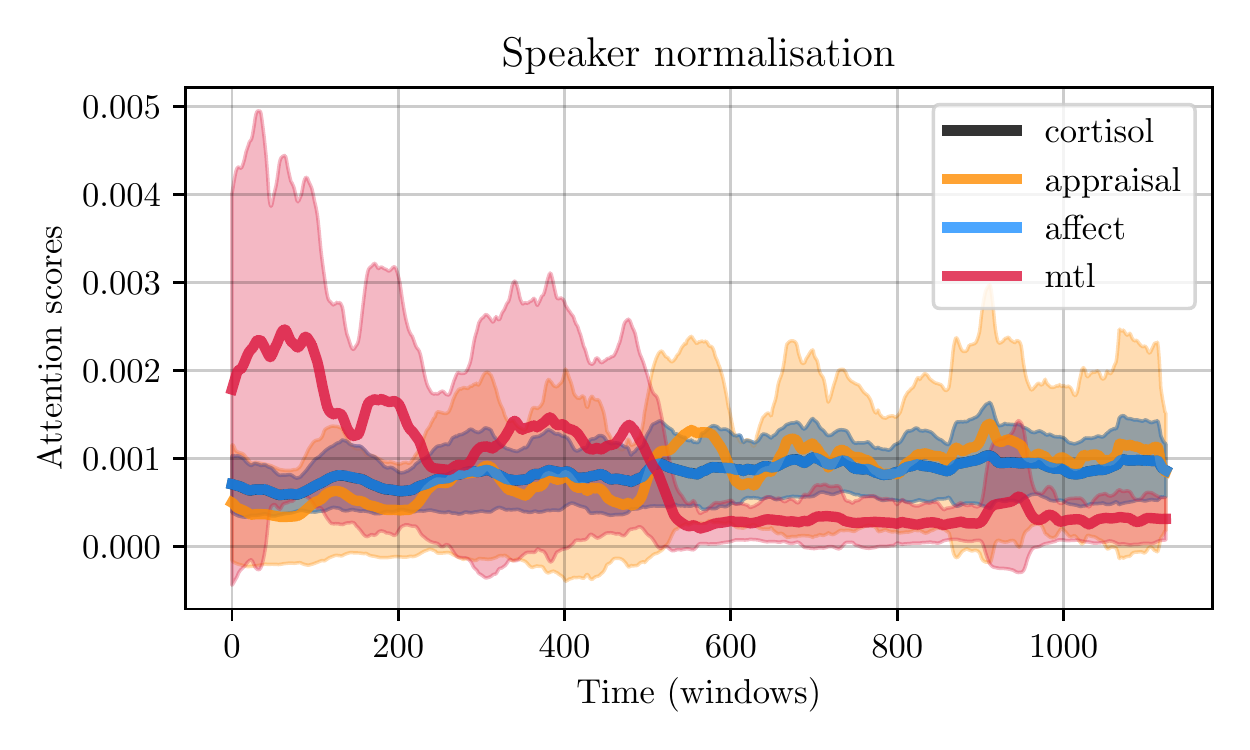}
    \end{subfigure}
    \caption{
    Attention weights obtained for the STL and MTL models using standard and speaker-based normalisation.
    We computed the mean (solid line) and standard deviation (shaded area) over all test subjects.
    Scores smoothed with a moving average window for better visibility.
    Cortisol and affect show very similar values and cannot be distinguished visually.
    }
    \label{fig:attention}
\end{figure*}

\Ac{MAE} results for all experiments are presented in \cref{table:results}.
Following standard practice, we evaluate all configurations on the validation set, but only the best ones on the test set to avoid overfitting.
Overall, the best results are obtained using attentional pooling and multitasking -- in line with previous results showing that both mechanisms generally improve performance.
This is true when considering the two factors in isolation as well: (A)\ac{GRU}-MTL is better than (A)\ac{GRU}-STL and A\ac{GRU} better than \ac{GRU}.
Moreover, \cref{table:results} shows that cortisol prediction yields consistently lower \ac{MAE} results than the other two targets.
While a comparison of (non-linear) regression results for different targets is not straightforward, this indicates that physiological responses to stress are more clearly manifested in the speech signal.

Interestingly, speaker-based normalisation yields the best performance for changes to cortisol and \ac{SI}, indicating that the effect of stress on voice differs across individuals.
Moreover, speaker-based normalisation is almost universally better than standard normalisation for all experiment configurations, lending further evidence to that claim.
This shows that more advanced personalisation approaches would be beneficial.

Finally, we visualised attention weights for all test samples in \cref{fig:attention}.
We computed the mean (solid line) and standard deviation (shaded area) over all subjects of the attention weights given by \cref{eq:alpha}.
This shows the importance assigned by the network to each timestep.
An interesting trend is revealed: the MTL network emphasises the first $\sim500$ samples of the \ac{TSST} -- a time span corresponding to the interview task.
This trend is particularly pronounced when using speaker-based normalisation.
In contrast, the STL network trained on the appraisal target shows a slight tendency towards the arithmetic task.
This shows how stress may manifest differently in the two \ac{TSST} tasks; future work could thus treat them separately. 

%%%%%%%%%%%%%%%%%%%%%%%%%%%%%%%%%%%%%%%%%%%%%%%%%%%%%%%%%%%%%%
\section{Conclusion}
\label{sec:conclusion}

We investigated the prediction of three stress-related constructs (changes to cortisol values, cognitive appraisal, and self-assessed affect) using speech features acquired during a \ac{TSST}.
Our analysis showed that the three constructs manifest differently in the speech signal, that there appear to be subject-specific effects in their manifestation, and that their information can be utilised in complimentary fashion.
Predicting those targets using speech signals obtained in a \ac{TSST} setting constitutes a first step towards the unobtrusive detection and characterisation of real-life stress episodes.

\addtolength{\textheight}{-12cm}   % This command serves to balance the column lengths

%%%%%%%%%%%%%%%%%%%%%%%%%%%%%%%%%%%%%%%%%%%%%%%%%%%%%%%%%%%%%%%%%%%%%%%%%%%%%%%%

%%%%%%%%%%%%%%%%%%%%%%%%%%%%%%%%%%%%%%%%%%%%%%%%%%%%%%%%%%%%%%%%%%%%%%%%%%%%%%%%

%%%%%%%%%%%%%%%%%%%%%%%%%%%%%%%%%%%%%%%%%%%%%%%%%%%%%%%%%%%%%%%%%%%%%%%%%%%%%%%%

\section*{ACKNOWLEDGMENT}

The work leading to this publication has received funding from the DFG's Reinhart Koselleck project No.\ 442218748 (AUDI0NOMOUS).
Data collection was supported by grant number DFG-KU 1401/6-1 assigned to B.\,M.\,K.

% \bibliographystyle{IEEEtran}
% \bibliography{bibliography.bib}

\section{\refname}
\printbibliography[heading=none]

\end{document}